\documentclass[a4paper,11pt]{article}
\usepackage{jinstpub} % for details on the use of the package, please see the JINST-author-manual
\usepackage{lineno}
\title{\boldmath Kinematic Reconstruction of Top-Quark Pairs Using Analytical and Machine-Learning Methods with CMS Open Data}

\author[a,b]{V.~Aushev,}
\author[a]{M.~Spitsyn,}
\author[a]{R.~Yermolenko,}
\author[c]{O.~Zenaiev}

\affiliation[a]{Taras Shevchenko National University of Kyiv, Kyiv, Ukraine}
\affiliation[b]{Institute for Nuclear Research, National Academy of Sciences of Ukraine, Kyiv, Ukraine}
\affiliation[c]{II.~Institute for Theoretical Physics, Hamburg University, Hamburg, Germany}

\emailAdd{oleksandr.zenaiev@desy.de}

\abstract{
The reconstruction of the top-quark-pair ($t\bar{t}$) kinematics in the dilepton $t\bar{t}$ decay channel is challenging due to the presence of two undetected neutrinos. In this work, existing kinematic reconstruction methods used by the CMS Collaboration are studied using a Monte Carlo sample with $t\bar{t}$ events from CMS Open Data at a centre-of-mass energy of 7 TeV. An improved version of the loose kinematic reconstruction, LKRv2, is developed by imposing more stringent requirements on the jet–lepton pairing entering the reconstruction. A further modification, LKRv3, exploits the kinematic similarity between the dilepton and neutrino-pair systems and improves the resolution of the reconstructed $t\bar{t}$ invariant mass relative to the baseline LKR by 7\%, with smaller improvements for the transverse momentum and rapidity of $t\bar{t}$. In addition, two machine-learning (ML) based approaches are developed using a residual multilayer perceptron and a conditional normalizing flow. 
Relative to LKRv3, they improve the resolution of the invariant mass, transverse momentum and rapidity by approximately 30\%, 20\% and 15\%, respectively. In particular, a resolution of 30 GeV is achieved for the reconstructed invariant mass of \tt near the production threshold. A dedicated generator-level study and a generator-to-detector transfer test demonstrate that the ML approaches learn a genuine physics-driven correction to the $\nu\bar{\nu}$ approximation, rather than merely exploiting detector-specific effects. These results demonstrate the strong potential of ML-based approaches for improving the reconstruction of dilepton $t\bar{t}$ events, while identifying the neutrino-pair approximation as an important limitation on the achievable resolution.
}

\keywords{Analysis and statistical methods, Data processing methods}

\arxivnumber{AAAA.BBBBB} % Only if you have one

\usepackage{color}
\usepackage{amsmath}
\usepackage{xspace}
\newcommand*{\ttbar}{\ensuremath{t\bar{t}}\xspace}
\renewcommand*{\tt}{\ttbar}
\newcommand*{\nunubar}{\ensuremath{\nu\bar{\nu}}\xspace}
\renewcommand*{\ll}{\ensuremath{l^{+} l^{-}}\xspace}
\newcommand*{\lplm}{\ensuremath{l^+l^-}\xspace}

\begin{document}

\maketitle
\flushbottom

\section{Introduction}
\label{sec:intro}
The top quark is the heaviest known elementary particle~\cite{ParticleDataGroup:2026mpi,Hoang:2020iah}. Due to its large mass the top quark plays a significant role in the Standard Model (SM) of particle physics, as well as in theories beyond the SM. Furthermore, the value of the top-quark mass, $m_t$, is important in cosmology in the context of the stability of the electroweak vacuum~\cite{Degrassi:2012ry,Alekhin:2012py}.

The study of the top quark is one of the key areas in the physics program at the Large Hadron Collider (LHC) where top quarks are produced copiously in $pp$ collisions, predominantly as pairs of top quark and antiquark (\ttbar) via the strong interaction. The top quark has a very short lifetime of approximately $5 \times 10^{-25}$ s and thus decays via the weak force before hadronization (which is driven by non-perturbative strong interactions and would happen at the time scale of $\sim 10^{-23}$s). 
As a result, its spin information is transferred to its decay products, providing direct access to the spin-related properties of the top quark~\cite{Mahlon:2010gw}.
On the other hand, to measure kinematic properties of top quarks (such as e.g.\ differential production cross sections) in experiments, the four-momenta of the top quark and antiquark need to be recovered from their decay products, which must first be identified and reconstructed in the detector and distinguished from other objects produced in the same $pp$ collision.

In the SM, almost always top quarks decay into a $W$ boson and a $b$ quark, owing to the large value of the corresponding element of the Cabibbo-Kobayashi-Maskawa (CKM) matrix $V_{tb} = 0.998$~\cite{ParticleDataGroup:2026mpi}.
The final state of the event with \ttbar can be classified by the further decays of the $W$ bosons. Three final states are distinguished: (1) in the dilepton decay channel ($10.5\%$) both $W$ bosons decay leptonically into a charged lepton and neutrino, (2) in the semileptonic decay channel ($43.8\%$) one $W$ boson decays leptonically while the other $W$ boson decays into a $q\bar{q}$ pair, (3) in the fully hadronic decay channel ($45.7\%$) both $W$ bosons decay into $q\bar{q}$ pairs. The $q\bar{q}$ pairs develop into two collimated bunches of particles which are called hadronic jets and are reconstructed in the detector as localized energy deposits. The leptonic decays of $W^{\pm}$ result in an isolated charged lepton and neutrino. Electrons and muons are easily reconstructed with very good resolution, while tau leptons are more difficult to reconstruct experimentally, therefore usually they are included in the experimental uncertainties via their leptonic decays~\cite{CMS:2024irj}. On the other hand, neutrinos escape undetected always.

Although the largest fraction of \ttbar events decay via the fully hadronic channel, those events are difficult to distinguish from the large QCD multijet background. The dilepton decay channel has the lowest branching fraction, but owing to the presence of two isolated charged leptons with large transverse momenta those events can be selected with high purity. In particular, events with $e^\pm\mu^\mp$ in the final state have a very low background contamination, making this channel particularly suitable for measurements of the total \ttbar production cross section~\cite{ATLAS:2023slx,CMS:2018fks}. The semileptonic decay channel has intermediate properties.

Measurements of differential \ttbar production cross sections require \ttbar kinematic reconstruction, i.e.\ the four-momenta of the top quark and antiquark must be recovered using their decay products. While this is rather straightforward in the fully hadronic decay channel, there is an unknown part associated with the neutrino(s) in the semileptonic and dilepton decay channels. In the dilepton \ttbar decay channel kinematic reconstruction is especially challenging due to the presence of two undetected neutrinos.

In this paper, existing \ttbar kinematic reconstruction methods (Section~\ref{sec:kinreco}) are tested on a publicly available CMS Monte Carlo (MC) data set using an analysis procedure similar to Refs.~\cite{CMS:2012hkm,CMS:2016yys} (Section~\ref{sec:evsel}). Then improvements to the reconstruction procedure are developed in Section~\ref{sec:lkrv3}, and new approaches based on machine learning (ML) are described in detail in Section~\ref{sec:nn}. Finally, conclusions are drawn in Section~\ref{sec:summary}.

%%%%%%%%%%%%%%%%%%%%%%%%%%%%%%%%%%%%%%%%%%%%%%%%%%%%%%%%%%%%%%%%%%%%%%
\section{Kinematic reconstruction in the dilepton \ttbar decay channel}
\label{sec:kinreco}
Several kinematic reconstruction techniques have been developed for dilepton \ttbar events and used in CMS measurements. They are described in detail in Ref.~\cite{CMS:2024irj}. Here the two algorithms are outlined: full and loose kinematic reconstruction. Only those features which are relevant for the further improvement are described in detail.

In {\bf full kinematic reconstruction (FKR)}, the three-momenta of the two neutrinos, corresponding to six unknowns, are recovered using the imbalance in the transverse momentum in the event, i.e.\ the so called missing transverse energy (MET, or $\vec{p}_T^{\textrm{miss}}$) 
(two constraints), and the known masses of the intermediate particles: $t$, $\bar{t}$, $W^+$ and $W^-$ (four constraints). Combined with the measured values of the lepton and jet three-momenta, this leads to a system of six equations. It is solved analytically, but it can have up to four solutions~\cite{Sonnenschein:2005ed,Sonnenschein:2006ud}. In the case of multiple solutions, the one with the lowest invariant mass of \ttbar is taken. Additional QCD radiation typically leads to the presence of more than two jets in the final state, thus further challenging the reconstruction algorithm. To deal with this, additional requirements on the invariant mass of the lepton ($l$) and jet ($b$) system are applied: $M(lb)<180$ GeV. This value follows from the inequality $M(lb) \le M(lb\nu) = m_t \approx 172$ GeV and allows for detector resolution and reconstruction effects. Additionally, jets are ranked according to their $b$-tag values, i.e.\ those which are likely initiated by $b$ quarks are preferred. Even with correctly selected input jets, detector effects may lead to a situation when no solution is found due to fluctuations in the measured values of the lepton and/or jet kinematic properties. To deal with this, the system of equations is solved multiple times with the input quantities smeared randomly according to their expected experimental resolution, and the weighted average of the solutions is taken using the expected $M(lb)$ spectrum for the weights~\cite{Korol:2016wzq}. The FKR finds a solution for about $90\%$ of \ttbar events. It became the standard algorithm used by CMS for \ttbar differential cross section measurements~\cite{CMS:2012hkm,CMS:2017iqf,CMS:2019esx,CMS:2024ybg} in the dilepton \ttbar decay channel.

Although it is highly efficient and robust, the FKR requires the $m_t$ value to be fixed a priori, and thus it is not well suited for measurements which are intended to be used for its extraction. For such cases, an alternative {\bf loose kinematic reconstruction (LKR)} algorithm was developed in Ref.~\cite{CMS:2019esx}. In this algorithm, only the \ttbar system as a whole is reconstructed, rather than the individual top quark and antiquark. This requires to find only four unknowns: the components of the four-vector of $\nu\bar{\nu}$. As in the FKR, the two constraints on the missing transverse energy are used to find the transverse momentum of $\nu\bar{\nu}$. The remaining two constraints are chosen to be equalities of the longitudinal momentum and energy of the charged leptons and neutrinos, i.e.\ $p_z(\lplm)=p_z(\nu\bar{\nu})$ and $E(\lplm)=E(\nu\bar{\nu})$. This exploits the fact that  both \nunubar and \lplm have negligible masses and both originate from the decays of $W$ bosons, thus their kinematic distributions must be the same. In a fraction of events, these constraints lead to unphysical values of $M(\nu\bar{\nu})<0$ or $M(\lplm\nu\bar{\nu})<2m_W$. In such cases they are replaced by physical constraints $M(\nu\bar{\nu})=0$ or $M(l^+l^-\nu\bar{\nu})=2m_W$. The selection of input jets is done in the same way as in the FKR. Owing to the simple relations, there is one solution always, and this method yields a similar reconstruction efficiency to the FKR.

ML approaches have also recently been investigated by the CMS Collaboration for the kinematic reconstruction of dilepton \ttbar events~\cite{CMSNOTE2025010}. Also an ML-based approach was used for the measurement of the differential cross section as a function of the invariant mass of \tt and jet, $M(\tt j)$, in Ref.~\cite{CMS:2022emx}.
In another study~\cite{Celotto2023} an ML-based kinematic reconstruction has been developed for $M(\tt)$ solely.
In contrast to Ref.~\cite{CMSNOTE2025010} where only a multilayer perceptron architecture was explored, in this work also a conditional normalizing flow model is considered, and in contrast to Refs.~\cite{CMS:2022emx,Celotto2023} the approach of this work is able to reconstruct the four-momenta of \tt rather than a single variable. 
A special emphasis is put on the reconstruction of the low $M(\ttbar)$ values, because the $M(\tt)$ distribution is mostly sensitive to the top-quark mass near the production threshold at $M(\tt) \approx 2 m_t$.
To our knowledge, this is the first work which uses CMS Open Data for ML studies of \tt kinematic reconstruction, conducted entirely outside the CMS Collaboration.

The kinematic reconstruction of dilepton $\ttbar$ events is affected by several distinct issues:
\begin{itemize}
    \item[(1)] the presence of two undetected neutrinos leaves their momenta unknown and requires additional kinematic relations to obtain a solution,
    \item[(2)] the presence of additional jets from QCD radiation introduces an ambiguity in assigning the reconstructed jets to the decay of \ttbar,
    \item[(3)] the finite experimental resolution (detector specific) of the reconstructed jets, MET and (to a lesser extent) leptons distort the kinematic inputs and thus the reconstructed \ttbar kinematics as well.
\end{itemize}
In this work, issue (1) is primarily addressed, while issue (2) is tackled to a lesser extent. However, all three issues are intertwined in detector-level reconstruction. In contrast, at generator level once the \ttbar system is produced and decayed, only issue (1) exists. In order to partially disentangle various effects, assess their individual contributions to the uncertainty of reconstructed \tt, and better understand the improvement provided by the new algorithms, the methods are applied not only at detector level (as customary in such studies), but also at generator level, where only issue (1) is present. 

%%%%%%%%%%%%%%%%%%%%%%%%%%%%%%%%%%%%%%%%%%%%%%%%%%%%%%%%%%%%%%%%%%%%%%
\section{Data sets and event selection}
\label{sec:evsel}
The CMS Collaboration has made a substantial amount of collision data from the LHC publicly available through the CERN Open Data Portal~\cite{cmsopendataportal}. The released data include fully reconstructed collision events collected with the CMS detector, together with corresponding detector-simulated MC samples and the software and documentation required for their analysis. These data provide a rich set of opportunities to reproduce and extend selected studies originally performed by the CMS Collaboration, as well as to develop and test new analysis methods outside the experimental collaboration.

In this paper, the detector-simulated MC sample \texttt{TTJets\_TuneZ2\_7TeV-madgraph-tauola} for 2011 collision data at a centre-of-mass energy $\sqrt{s}=7$ TeV is used~\cite{cmsopendataset}. It contains \ttbar events generated using the \texttt{MadGraph} event generator (version 5.1.1.0)~\cite{Alwall:2014hca}. 
The generated events were subsequently processed with PYTHIA (version 6.424)~\cite{Sjostrand:2006za} for parton showering and hadronization, and further with the \texttt{Geant}-based simulation of the CMS detector.
No real $pp$ collision data are used in this work.

MC events in the AODSIM format are reconstructed using the CMS software and further filtered using the following requirements on \ttbar decay products:
\begin{itemize}
    \item there must be at least two leptons with transverse momentum $p_T > 20$ GeV, pseudorapidity $|\eta|<2.4$ and opposite electric charges. These leptons must be isolated using the criterion $I_{rel}^{\Delta R<0.3}<0.17(0.20)$ for electrons (muons), where $I_{rel}^{\Delta R<0.3}$ is defined as the $p_T$ sum of all neutral hadrons, charged hadrons, and photon candidates within a distance of $0.3$ from the lepton in $\eta-\phi$ space, divided by the $p_T$ of the lepton. The invariant mass of the leptons with the highest $p_T$ value is required to satisfy $M(ll)>12$ GeV. In the $ee$ and $\mu\mu$ decay channels, additional requirements $M(ll)<76$ GeV or $M(ll)>106$ GeV, and $\text{MET}>30$ GeV are applied to suppress background from events with Drell-Yan lepton pair production,
    \item  there must be at least two jets with $p_T > 30$ GeV and $|\eta|<2.4$ reconstructed by the anti-$k_T$ algorithm~\cite{Cacciari:2008gp} with a clustering parameter $R=0.5$. The jet energies are corrected using the corrections provided with the CMS software. Jets that likely originate from the hadronization of $b$ quarks are identified by the combined secondary-vertex (CSV) algorithm~\cite{CMS-PAS-BTV-11-004} and assigned the so called $b$-tagging value. At least one of the two jets must be $b$-tagged using the loose working point with the $b$-tagging value $0.244$.
\end{itemize}
This event selection procedure is similar to the one used by the CMS Collaboration in Refs.~\cite{CMS:2012hkm,CMS:2016yys}. It is realized using the public code example provided by CMS~\cite{github-ttbar2011}, where additional trigger requirements as well as requirements on the quality of object and event reconstruction are implemented. More details can be found in Ref.~\cite{CMS:2012hkm}. Since the sample of events is a clean \ttbar signal sample, these further selection requirements have only minor impact.
After event selection, the kinematic reconstruction of \tt is performed using the LKR method described in Section~\ref{sec:kinreco} as well as newly developed methods described in Sections~\ref{sec:lkrv3} and \ref{sec:nn}. All three dilepton decay channels ($ee$, $\mu\mu$, $e\mu$) are combined.

%%%%%%%%%%%%%%%%%%%%%%%%%%%%%%%%%%%%%%%%%%%%%%%%%%%%%%%%%%%%%%%%%%%%%%
\section{Improved loose kinematic reconstruction}
\label{sec:lkrv3}
As explained in Section~\ref{sec:kinreco}, the LKR algorithm does not reconstruct individual $\nu$ and $\bar{\nu}$ momenta, but only their sum \nunubar. Therefore it does not use any assumption on which of the two jets and charged leptons originate from the decay of the top quark or antiquark.
Only jets that satisfy the $M(lb)<180$ GeV requirement are used in the kinematic reconstruction. In the original LKR, this includes the situation when only one jet is successfully paired with each of the leptons and satisfy $M(lb)<180$ GeV, while the other jet does not satisfy this criterion if paired with either of the two leptons.
The first proposed improvement of the LKR modifies the jet selection procedure such that both jets must fulfill the $M(lb)<180$ GeV requirement if paired with either of the two leptons, and analogously both leptons must satisfy this requirement if paired with either of the two jets. This modification of the LKR algorithm is further referred to as {\bf LKRv2}.

\begin{figure}
	\centering
	\includegraphics[width=1.0\textwidth]{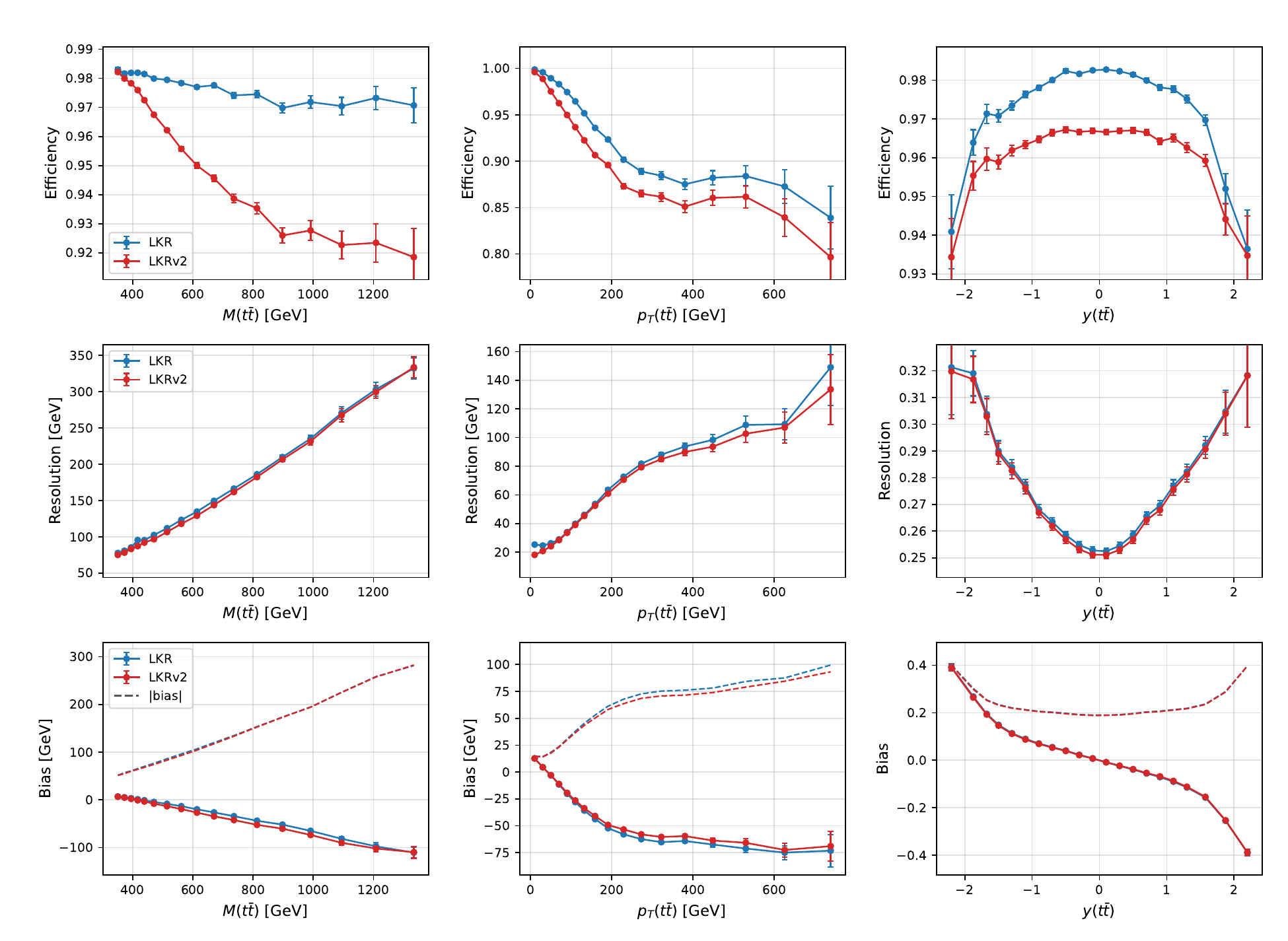}
	\caption{The efficiency (upper row), resolution (middle row) and bias (lower row)
	of the LKR and LKRv2 algorithms as a function of the generated $M(\ttbar)$ (left),
	$p_T(\tt)$ (middle) and $y(\tt)$ (right).}
	\label{fig:lkr_lkrv2}
\end{figure}

In Fig.~\ref{fig:lkr_lkrv2} performance of the LKR and LKRv2 algorithms is presented in terms of reconstruction efficiency, resolution and bias. The efficiency of each method is defined as the ratio of the number of events which pass the kinematic reconstruction to the total number of events after applying all selection criteria described in Section~\ref{sec:evsel}. The resolution is defined as the RMS of the difference between the reconstructed and the generated quantity, and the bias as its mean. Apart from the mean of the difference, the mean of the absolute difference (|bias|) is reported as well. In the latter metric, biases of
opposite sign in different events do not cancel.
The performance of the LKR is similar to the CMS results published in Ref.~\cite{CMS:2024irj}, however, one should notice that the latter were obtained at different center-of-mass $\sqrt{s}=13$ TeV and using different event selection criteria.
Compared to the LKR results obtained in this work, the LKRv2 yields resolution improvement by a few $\%$. This is accompanied by a marginal drop of efficiency of about $1\%$, which is expected since the LKRv2 features more stringent requirements on the input jets.

The original LKR exploits the kinematic similarity of the \nunubar and \ll systems by assigning the unknown longitudinal momentum and energy of the $\nu\bar{\nu}$ system to the ones of the \ll system. As mentioned in Section~\ref{sec:kinreco}, in some events this leads to unphysical negative values $M(\nunubar)<0$ and/or $M(\nunubar\lplm)-2m_W<0$ which must be corrected by explicitly setting their values to the minimal physical values, i.e.\ to zero.%
\footnote{The small masses of leptons and neutrinos are neglected.}
An alternative variant of the LKR proceeds by assigning the invariant mass $M(\nunubar)=M(\lplm)$ and rapidity $y(\nunubar)=y(\lplm)$. This modification is applied on top of the LKRv2 algorithm and is further referred to as {\bf LKRv3}.

\begin{figure}
	\centering
	\includegraphics[width=1.0\textwidth]{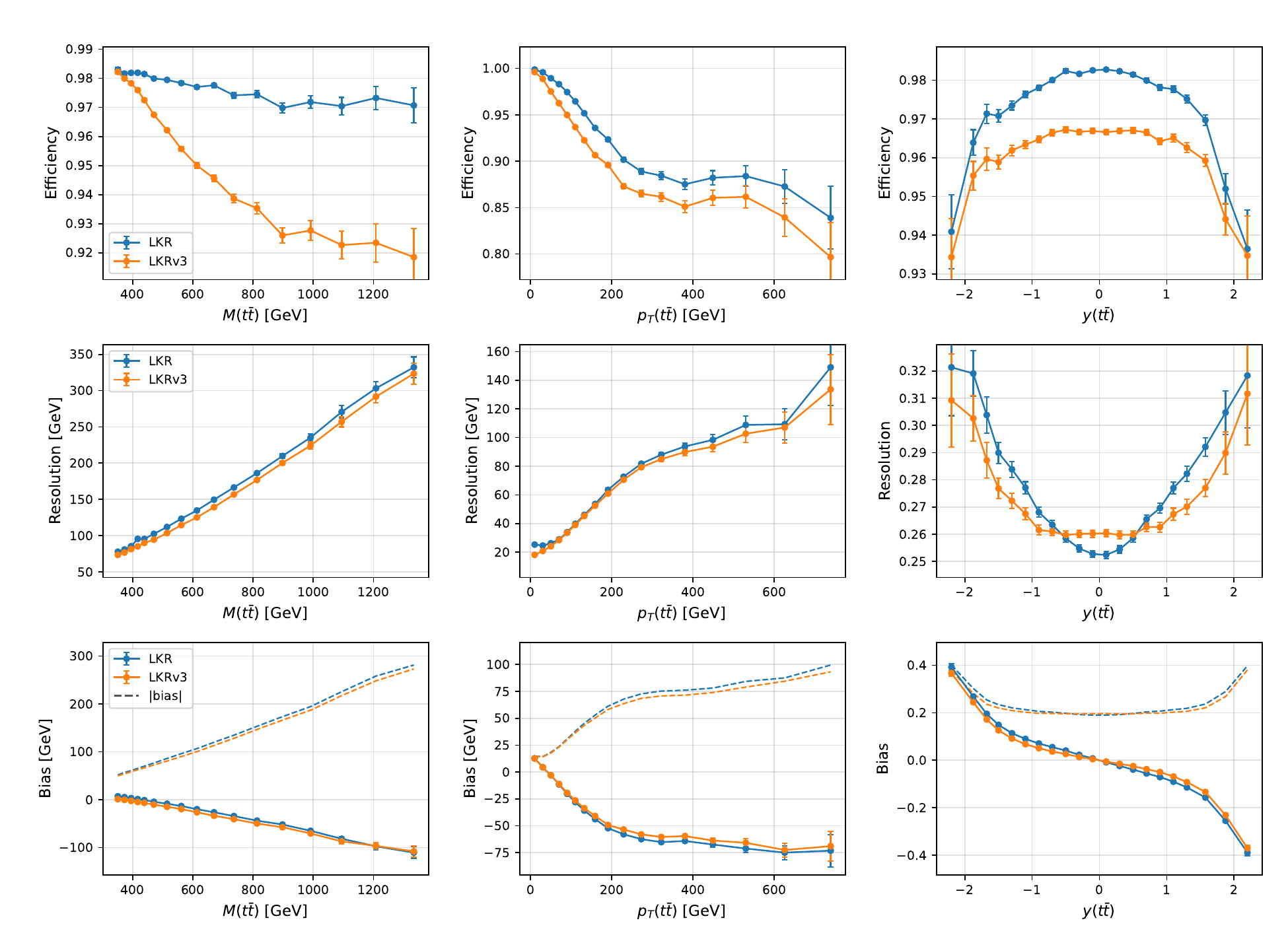}
	\caption{Same as Fig.~\ref{fig:lkr_lkrv2}, but comparing the LKR and LKRv3 algorithms.}
	\label{fig:lkr_lkrv3}
\end{figure}

In Fig.~\ref{fig:lkr_lkrv3} the efficiency, resolution and bias of the LKR and LKRv3 algorithms is compared.
The LKRv3 improves the resolution of $M(\ttbar)$ substantially. However, as a drawback the resolution of $y(\tt)$ gets slightly worse around $y(\tt)\approx 0$.
Since LKRv3 modifies only the longitudinal components of the $\nunubar$ system,
its $p_T(\tt)$ distribution is identical to that of LKRv2.
As in the LKRv2, there is a marginal drop of efficiency of the LKRv3 due to the more stringent input jet requirements. The corresponding effect on non-\ttbar background events is not studied in this work. We expect these requirements to suppress background events as well, potentially resulting in an overall improvement in signal significance despite the small loss in signal efficiency.%
\footnote{The signal significance is typically proportional to $N_s/\sqrt{N_s+N_{bg}}$ where $N_{s}$ and $N_{bg}$ are the numbers of signal and background events, respectively. An ideal reconstruction algorithm would have 100\% efficiency for signal events and 0\% efficiency for background events. In practice, however, an algorithm with significantly lower efficiency for background events than for signal events might be preferred over a kinematic reconstruction algorithm which has a solution always.}

%%%%%%%%%%%%%%%%%%%%%%%%%%%%%%%%%%%%%%%%%%%%%%%%%%%%%%%%%%%%%%%%%%%%%%
\section{NN kinematic reconstruction}
\label{sec:nn}

The deviation of the true \nunubar kinematics from the LKRv3 approximation
(Section~\ref{sec:lkrv3}) is not random but correlates with the kinematics of the observed decay
products. A neural network (NN) can learn this systematic part. The NN predicts
$\ln M(\tt)$, $\ln(p_T(\tt)+1~\mathrm{GeV})$ and $y(\tt)$ directly from the observed objects (direct
regression). The logarithmic form makes the errors in $M$ and $p_T$ relative over the whole spectrum,
and the $1$~GeV offset regularizes events with small $p_T(\tt)$. Two models are used: a multilayer
perceptron ({\bf MLP}) and a conditional normalizing flow ({\bf Flow}), which predicts the full
per-event density $p(M(\tt), p_T(\tt), y(\tt)\,|\,\text{inputs})$.

\subsection{Architecture, inputs and training}
\label{sec:nn_setup}

The inputs and the targets are standardized to zero mean and unit variance, with the means and
variances computed on the training events and stored together with the network. The MLP consists of
a linear layer of width $128$ with layer normalization~\cite{Ba:2016nlm} and SiLU activation
function~\cite{Elfwing:2017aug}, three residual blocks~\cite{He:2015wrn} and a linear output layer,
in total $1.0\times10^5$ parameters. Each residual block transforms the hidden vector $h$ as
$h \to \mathrm{SiLU}\big(h + \mathrm{LN}(W_2\,\mathrm{SiLU}(\mathrm{LN}(W_1 h)))\big)$, where $W_{1,2}$
are linear layers and LN is the layer normalization. The MLP is trained with the L1 loss (mean absolute
error), as it estimates the median of the targets for given inputs and gives a smaller bias and a better
conditioned response than the L2 loss, as explained further in Section~\ref{sec:nn_arch}.The loss is the average over the three standardized targets with equal weights, so that none of the variables is preferred.

The Flow is a neural spline flow~\cite{Durkan:2019nsf}: a three-dimensional standard normal
distribution is transformed into the distribution of the targets by four autoregressive
rational-quadratic spline transformations with 8 bins each. The parameters of the splines are
computed by small networks (two hidden layers of width 128) from the previous target components and
from a 64-dimensional context vector, which is computed from the inputs by the same residual network
as in the MLP; in total the Flow has $3.8\times10^5$ parameters. It is trained by minimizing the
negative log-likelihood of the true values. The point estimate is the mean of 64 samples per event,
which are generated from the same fixed set of quasi-random (Sobol) points for all events, which makes the
estimate deterministic.

The inputs of the neural networks are based on 24 variables: the four-momenta of $l^-$, $l^+$ and of the two jets
selected by LKRv3 and the missing transverse momentum (18 variables, ``raw''), additional four lepton--jet masses $M(l^\pm j_{1,2})$ and two CSV $b$-tagging discriminants of the two jets. 
The selection of input jets is done in the same way as in the LKRv2/LKRv3, therefore the NN methods have the same reconstruction efficiency of $96.5\%$.
All inputs are reconstructed quantities, so no generator-level information enters the
networks at detector level. Additional jets are not used, since their modeling in the MC may be unreliable. 

The $730 \times 10^3$ events in which LKRv3 finds a suitable jet pair are split randomly with a fixed seed
into $70\%$ training, $15\%$ validation and $15\%$ test events. The same split is used for all networks. Both models are trained with the AdamW
optimizer~\cite{Loshchilov:2017adw} with a learning rate of $5\times10^{-4}$, decreasing to
$5\times10^{-6}$ with cosine annealing, weight decay $10^{-2}$ and batch size 4096, for at most 400
epochs. Early stopping is triggered when the loss on the validation events does not improve for 40
epochs, and the network of the epoch with the lowest validation loss is kept. The validation events
are used only for this choice, and the test events are not used at any stage of the training or of
the choice of the model. All results are obtained on the test events, and the same events are used for all
comparisons.
The training takes a few minutes for the MLP and up to an hour for the Flow on a single
GPU. Repeating the training with five different seeds changes the RMS of $M(\tt)$ by $0.03$--$0.1$~GeV
and the condition number of the response matrix (Section~\ref{sec:nn_metrics}) by $0.1$.

To improve generalization and make explicit use of the azimuthal symmetry of $pp$
collisions, each training event is additionally augmented by a random rotation around
the beam axis ($z$-axis). For a given event, a single angle
$\phi \in [0, 2\pi)$ is drawn, and the transverse momentum components of both charged
leptons, both selected jets, and the missing transverse momentum are rotated
consistently:
\begin{equation}
  \begin{pmatrix} p_x' \\ p_y' \end{pmatrix}
  =
  \begin{pmatrix} \cos\phi & -\sin\phi \\ \sin\phi & \cos\phi \end{pmatrix}
  \begin{pmatrix} p_x \\ p_y \end{pmatrix} .
  \label{eq:rotation_augmentation}
\end{equation}
The longitudinal momentum and energy components of all objects are unaffected by this
transformation, as are the lepton--jet masses $M(l^\pm j_{1,2})$ and the CSV discriminants,
which are invariant under rotations around the beam axis. Since a rotated event corresponds to a
physically equivalent configuration with the same $M(t\bar t)$, $p_T(t\bar t)$ and
$y(t\bar t)$, this augmentation effectively increases the diversity of the training
sample without altering the regression targets. A new random angle is drawn
independently for every event in every training batch, so the network is exposed to a
continuously varying azimuthal orientation over the course of training.
In addition, the reflection $z\to-z$ is applied to half of the events in every batch: it flips the
sign of the longitudinal momenta of all objects and of the target $y(t\bar t)$, while $M(t\bar t)$ and
$p_T(t\bar t)$ are unchanged. Both transformations are symmetries of the production and, to a good
approximation, of the detector, and it was verified that they do not distort the MC distributions.
The augmentation is applied only during training and is not used for validation, testing,
or inference.

\subsection{Metrics}
\label{sec:nn_metrics}

The performance of the methods is evaluated in terms of efficiency, bias and resolution (see Section~\ref{sec:lkrv3}).
With about $10^5$ test events, the statistical
uncertainty of the resolution of $M(\tt)$ is about $0.2$~GeV. 
In addition to these quantities, the two further metrics described below
are related to cross section measurements as a function of $M(\tt)$ and their achievable precision.

{\it Direct folding.} The nominal MC (theory A) is compared with the same MC, but reweighted event by event
with $w(M^\mathrm{gen}(\tt))$ (theory B, normalized to the same number of events), in bins of
reconstructed $M(\tt)$:
\begin{equation}
  \Delta\chi^2=\sum_i \frac{(y_{B,i}-y_{A,i})^2}{y_{A,i}},
  \label{eq:fold}
\end{equation}
where $y_{A,i}$ and $y_{B,i}$ are the numbers of events in bin $i$. Three shapes of $w$ are used: a
slope $w=(M/500~\mathrm{GeV})^{0.1}$, a narrow resonance $1+0.5\,G(M;500,30~\mathrm{GeV})$ and a
`toponium'-like threshold excess $1+0.3\,G(M;350,5~\mathrm{GeV})$, where $G(x; M, \sigma)$ is a Gaussian distribution of $x$ with mean $M$ and width $\sigma$. Their
amplitudes are scaled so that an ideal detector (reconstructed $=$ generated) gives $\Delta\chi^2=25$,
which corresponds to a $5\sigma$ separation. The fraction of this value retained by a method measures how much of the
shape information survives the reconstruction. The reconstructed bins are common for all methods and
as narrow as the MC statistics allows: starting from the maximum of the spectrum, each bin is the
narrowest one containing at least 100 events for every method (up to three edge bins with at least 20 events), and the bin width never decreases
towards the tails. This avoids a choice of the bins
tuned to a particular method.

{\it Condition number.} The response matrix $A_{ij}$ is the probability that an event generated in
bin $j$ (generator level) is reconstructed in bin $i$ (detector level), including the efficiency. Its condition number
$\mathrm{cond}(A)=\sigma_1/\sigma_n$, the ratio of the largest to the smallest singular
value~\cite{Hocker:1995kbI}, is an upper bound on the amplification of relative uncertainties in
unfolding; $\mathrm{cond}(A)\approx1$ for a diagonal response. The bins at generator level are those of the
cross section measurement, $345|400|470|550|650|800|1100|\infty$~GeV, and each of them is split into
two bins at detector level.

\subsection{Results at detector level}
\label{sec:nn_results}

\begin{table}[htbp]
  \centering
  \small
  \begin{tabular}{lccccccccc}
    \hline\hline
    Method & Eff.\ [\%] & $|\mathrm{bias}|(M)/\textrm{GeV}$ & $\sigma(M)/\textrm{GeV}$ & $\sigma(p_T)/\textrm{GeV}$ & $\sigma(y)$ &
      \multicolumn{3}{c}{$\Delta\chi^2$} & cond$(A)$ \\
      & & & & & & slope & res. & thr. & \\
    \hline
    ideal             & 100  & 0    & 0     & 0    & 0     & 25.0 & 25.0 & 25.0 & 1.2 \\
    LKR               & 98.0 & 83.2 & 115.3 & 39.6 & 0.272 & 12.9 & 2.1  & 1.6  & 39.3 \\
    LKRv3             & 96.5 & 78.8 & 107.7 & 37.9 & 0.268 & 13.2 & 2.4  & 1.7  & 26.7 \\
    MLP               & 96.5 & 49.5 & 75.2  & 30.8 & 0.233 & 16.4 & 4.7  & 3.4  & 8.5 \\
    Flow              & 96.5 & 49.7 & 74.1  & 30.5 & 0.229 & 16.5 & 4.6  & 3.3  & 9.4 \\
    \hline\hline
  \end{tabular}
  \caption{Performance at detector level on the test events: efficiency, mean absolute deviation
  and resolution, $\Delta\chi^2$ of the folding test for the slope, resonance and
  threshold scenarios and condition number of the response matrix.}
  \label{tab:nn}
\end{table}

\begin{figure}[htbp]
  \centering
  \includegraphics[width=1.0\textwidth]{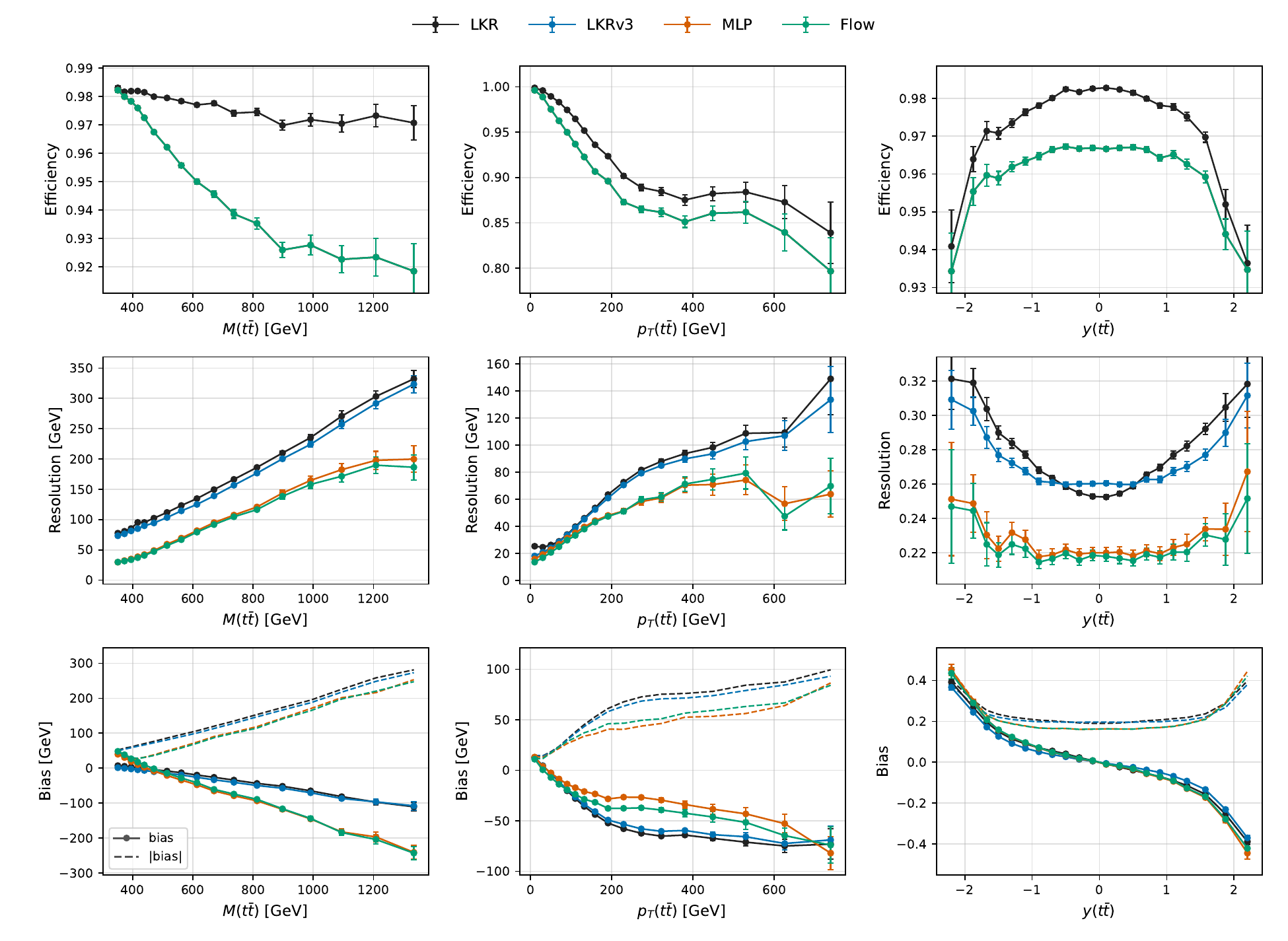}
  \caption{Efficiency, resolution and bias (dashed: mean absolute deviation) as a function of the
  generated $M(\tt)$, $p_T(\tt)$ and $y(\tt)$. The NN methods are evaluated on test events only.}
  \label{fig:nn_res}
\end{figure}

\begin{figure}[htbp]
  \centering
  \includegraphics[width=1.0\textwidth]{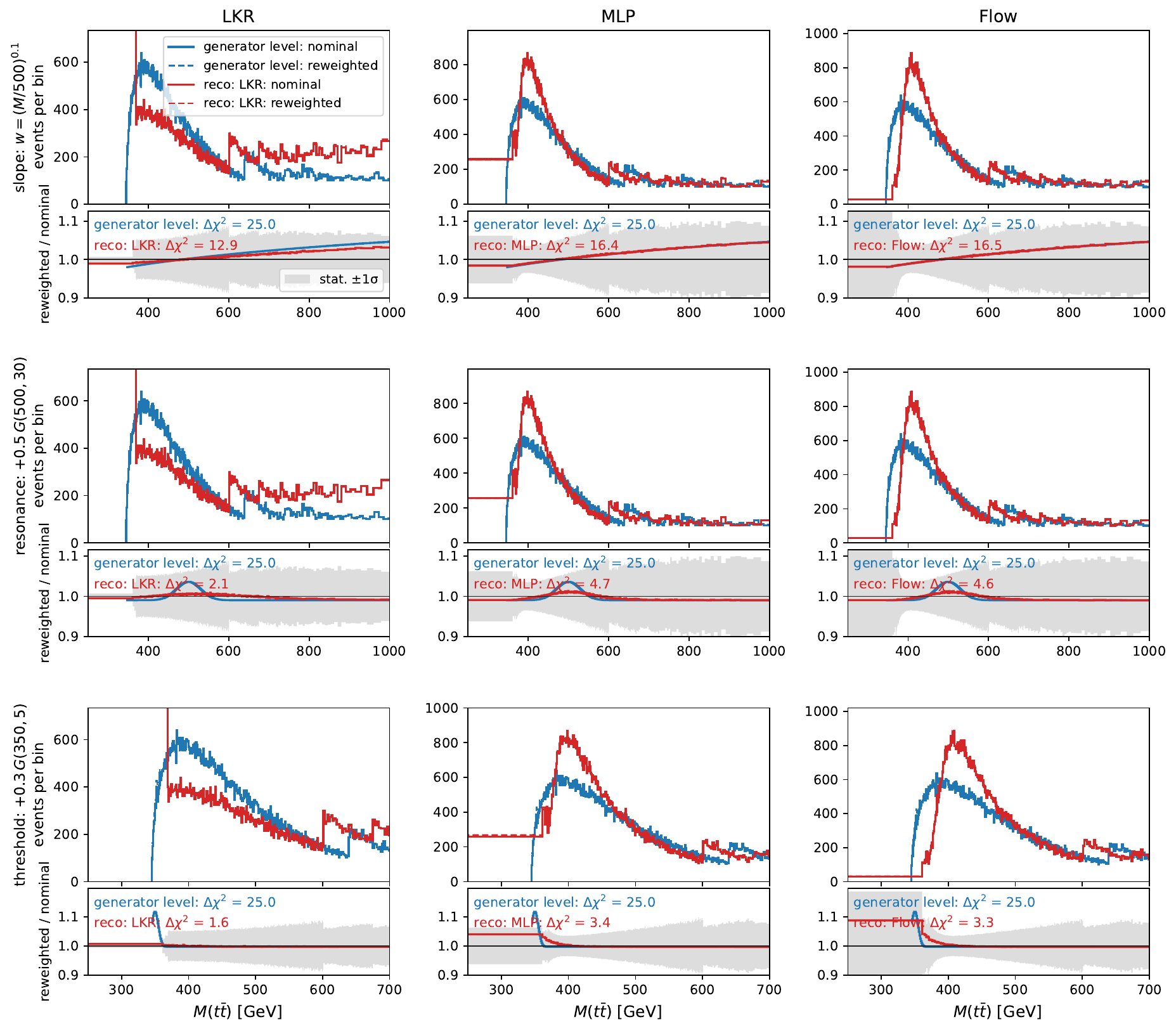}
  \caption{Direct folding test for the slope, resonance and threshold scenarios (rows) and LKR, MLP
  and Flow (columns). Upper panels: nominal (solid) and reweighted (dashed) $M(\tt)$ distributions at
  generator level (blue, fine bins) and after reconstruction (red, common reconstructed bins); lower
  panels: their ratio. $\Delta\chi^2$ is computed in the reconstructed bins.}
  \label{fig:nn_fold}
\end{figure}

\begin{figure}[htbp]
  \centering
  \includegraphics[width=1.0\textwidth]{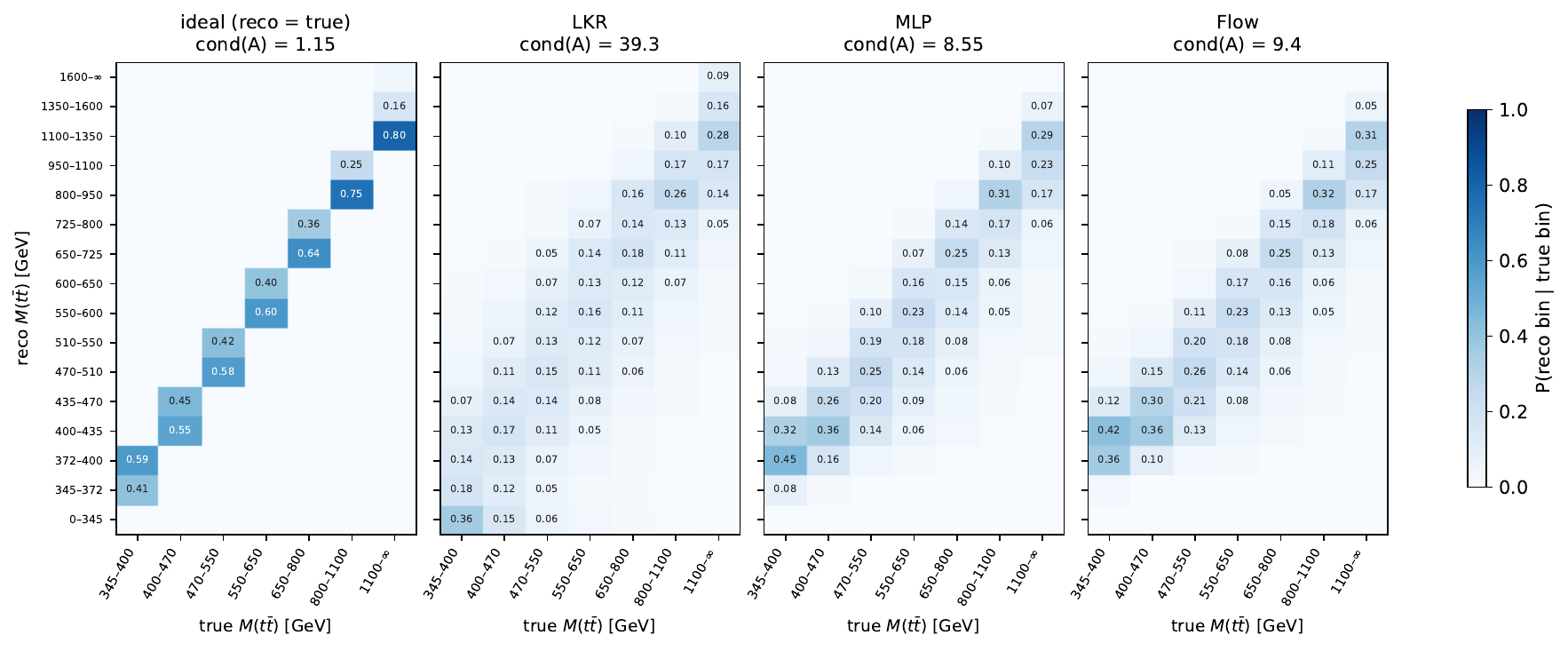}
  \caption{Response matrices $P(\text{reconstructed bin}\,|\,\text{generated bin})$ for $M(\tt)$ and
  their condition numbers.}
  \label{fig:nn_resp}
\end{figure}

The results are given in Table~\ref{tab:nn} and Figs.~\ref{fig:nn_res}--\ref{fig:nn_resp}. Both NN
methods improve $\sigma(M(\tt))$ by $35\%$ with respect to LKR and by $30\%$ with respect to LKRv3,
$\sigma(p_T(\tt))$ by $22\%$ and $\sigma(y(\tt))$ by $15\%$. These differences are much larger than
the statistical uncertainties and the spread between trainings. The improvement is present in all
bins of $M(\tt)$, $p_T(\tt)$ and $y(\tt)$ (see Fig.~\ref{fig:nn_res}) and is largest near the threshold,
where the best achieved $M(\tt)$ resolution is $30$~GeV compared to $78$~GeV for LKR, a factor of $2.6$ better. In
the tail of the spectrum the gain is about $1.5$. The MLP and the Flow perform the same within a few
per cent: the Flow is slightly better above $M(\tt)\approx500$~GeV, the MLP has a smaller bias at the
threshold.

The price of the higher resolution is an $M$-dependent bias towards the most populated region of the
spectrum ($M\approx450$~GeV): about $+40$~GeV at the threshold and $-100$~GeV at $M^\mathrm{gen}(\tt)\approx800$~GeV.
This is a property of any estimator minimizing the average deviation (``regression to the mean''):
if the inputs of an event are compatible with a range of true values, the estimate is shifted towards
the values that are more frequent in the training sample~\cite{Baattrup:2026inh}. The bias is accounted for by the response
matrix in folding and unfolding, and its possible dependence on the shape of the MC spectrum has to be
considered as a systematic uncertainty. Since the NN estimates are more precise event by event, the
mean absolute deviation is reduced from $83$ to $50$~GeV. For the Flow, the per-event
density is cut at $2m_t$ and has a long upper tail. As a result, its mean is shifted upwards near the
threshold more than the median estimated by the MLP.

In the folding test the NN methods retain about $66\%$ of the ideal $\Delta\chi^2$ for the slope,
compared to $52\%$ for LKR. They are about $1.3$ times more sensitive to a smooth shape change and
twice as sensitive to the narrow structures (resonance and threshold), where the gain from the better
resolution is largest. The gain is smaller than the ratio of the resolutions, because the regression
to the mean compresses the reconstructed spectrum. The condition number of the response matrix
decreases by a factor of 4--5, from $39$ (LKR) to $8.5$ (MLP) and $9.4$ (Flow). In unfolding, this
means a much smaller amplification of the uncertainties and a weaker regularization
(Fig.~\ref{fig:nn_resp}).

\subsection{Results at generator level}
\label{sec:gen}

To separate the neutrino approximation from the jet assignment and the detector resolution, the
networks are also trained and applied at generator level, where the inputs are the true leptons, the
$b$ quarks and the transverse momentum of the neutrino pair, without the CSV discriminants, which are
a detector feature. All $2.7\times10^6$ generated dilepton events are used, with the same 70/15/15
split; at this level all methods have an efficiency very close to $100\%$. 

\begin{table}[htbp]
  \centering
  \begin{tabular}{llcccc}
    \hline\hline
    Variable & Method & gen--gen & det--det (no CSV) & det--det & gen$\to$det \\
    \hline
    $\sigma(M)$ [GeV]   & LKRv3 & 90.1  & --    & 107.7 & 107.7 \\
                        & MLP   & 50.3  & 77.1  & 75.2  & 100.6 \\
                        & Flow  & 50.4  & 76.2  & 74.1  & 96.8  \\
    \hline
    $\sigma(p_T)$ [GeV] & LKRv3 & 8.6   & --    & 37.9  & 37.9  \\
                        & MLP   & 8.3   & 32.9  & 30.8  & 38.0  \\
                        & Flow  & 8.1   & 32.6  & 30.5  & 38.1  \\
    \hline
    $\sigma(y)$         & LKRv3 & 0.255 & --    & 0.268 & 0.268 \\
                        & MLP   & 0.158 & 0.243 & 0.233 & 0.289 \\
                        & Flow  & 0.157 & 0.239 & 0.229 & 0.283 \\
    \hline\hline
  \end{tabular}
  \caption{Resolution of $M(\tt)$, $p_T(\tt)$ and $y(\tt)$ for networks trained and applied at generator (gen) or detector
  (det) level; gen$\to$det: trained at generator level and applied to detector-level inputs without
  retraining. All numbers are obtained on test events.}
  \label{tab:nn_gen}
\end{table}

Table~\ref{tab:nn_gen} shows that even with perfect inputs LKRv3 leaves $\sigma(M(\tt))=90$~GeV,
while $\sigma(p_T(\tt))$ drops to $8.6$~GeV, since
$p_T(\tt)$ is limited by the MET resolution mainly. The networks reduce
$\sigma(M(\tt))$ to $50$~GeV, a factor of $1.8$, i.e.\ most of the limitation of the analytical algorithms
comes from the neutrino approximation, and its substantial part can be learned from the visible kinematics. The detector
degrades the NN resolution from $50$ to $75$~GeV. Networks trained at generator level and applied to
detector-level inputs are better than LKRv3 ($97$--$101$ vs.\ $108$~GeV) and do not depend on the
detector simulation, but they retain only part of the improvement of the detector-level
training, since the optimal estimate depends on the resolution of the inputs.

\subsection{Further studies and cross-checks}
\label{sec:nn_arch}

\begin{table}[htbp]
  \centering
  \begin{tabular}{llcccc}
    \hline\hline
    Method & Input & $\sigma(M)$ & $\sigma(M)_\mathrm{thr}$ & $\Delta\chi^2_\mathrm{thr}$ & cond$(A)$ \\
    \hline
    MLP & raw                 & 77.5 & 36.5 & 3.2 & 8.9 \\
        & raw + $M(lj)$       & 77.1 & 36.1 & 3.3 & 8.4 \\
        & raw + CSV           & 75.4 & 30.2 & 3.3 & 9.0 \\
        & raw + CSV + $M(lj)$ & 75.2 & 29.8 & 3.4 & 8.5 \\
    \hline
    Flow & raw                & 76.8 & 34.3 & 3.2 & 9.9 \\
         & raw + CSV + $M(lj)$& 74.1 & 29.9 & 3.3 & 9.4 \\
    \hline\hline
  \end{tabular}
  \caption{Detector-level results for different inputs.
  $\sigma(M)_\mathrm{thr}$ (GeV): $340<M^\mathrm{gen}(\tt)<362$~GeV.}
  \label{tab:inputs}
\end{table}

{\it Inputs} (Table~\ref{tab:inputs}). According to the performed tests, the 18 raw variables give most of the improvement. The four
$M(lj)$ masses improve $\sigma(M(\tt))$ by only $0.5$~GeV.
The CSV discriminants improve the resolution by 2~GeV over the whole spectrum and by $17\%$ near the threshold, where
the jets are soft and the choice of the $b$ jets matters most. Above $M(\tt)\approx550$~GeV they bring
no improvement. 

\begin{figure}[htbp]
  \centering
  \includegraphics[width=1.0\textwidth]{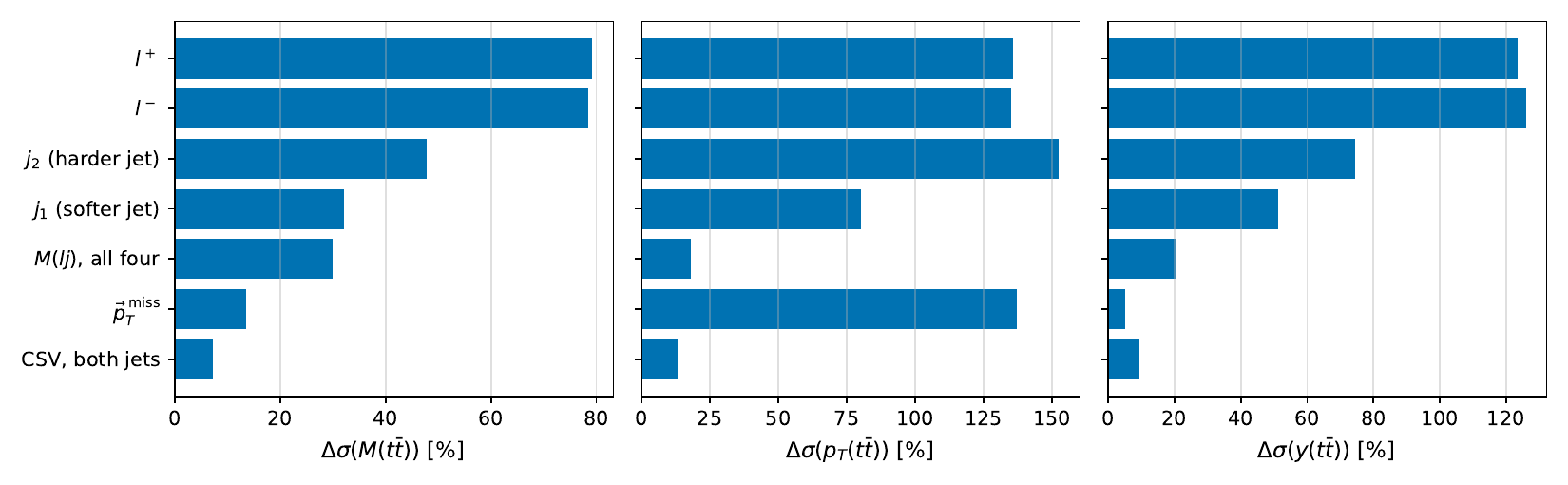}
  \caption{Permutation importance of the inputs of the MLP at detector level: relative increase of the
  resolution of $M(\tt)$, $p_T(\tt)$ and $y(\tt)$ when the inputs of one object are shuffled between
  the test events. 
  For the leptons and jets, all
  four components of the four-momentum are shuffled together.}
  \label{fig:perm}
\end{figure}

{\it Permutation importance.} In order to understand which inputs of the trained MLP are most important, 
the values of one input object are shuffled randomly between the test events, while all other inputs are kept, and the increase of the
resolution is measured without network retraining. The results are shown in Fig.~\ref{fig:perm}. When the four-momentum of a
lepton or a jet is shuffled, the $M(lj)$ masses are recomputed, so that the inputs remain consistent.
The four-momenta of the leptons are the most important inputs for $M(\tt)$ and $y(\tt)$: shuffling
either lepton worsens $\sigma(M(\tt))$ by about $80\%$ and $\sigma(y(\tt))$ by about $125\%$. The jets 
are the second most important inputs. The MET is important mainly for $p_T(\tt)$, in agreement with the generator-level
study (Section~\ref{sec:gen}). Shuffling the CSV discriminants changes the overall resolution by only
$7\%$, since their effect is concentrated near the threshold. The importance of the $M(lj)$ masses
($30\%$ for $M(\tt)$) and their small gain in Table~\ref{tab:inputs} are not contradictory: after the shuffling, they are inconsistent with the four-momenta, which the network does not expect, while without them the network extracts the same information from the four-momenta.

{\it Correction of the LKRv3 solution.} Instead of predicting $M(\tt)$, $p_T(\tt)$ and $y(\tt)$
directly, the same MLP was trained to predict corrections to the LKRv3 solution,
$\ln(M^\mathrm{gen}/M^\mathrm{LKRv3})$, $\ln((p_T^\mathrm{gen}+1~\mathrm{GeV})/(p_T^\mathrm{LKRv3}+1~\mathrm{GeV}))$
and $y^\mathrm{gen}-y^\mathrm{LKRv3}$, with and without limiting the corrections to $5\%$ for $M(\tt)$,
$20\%$ for $p_T(\tt)$ and $0.05$ for $y(\tt)$. Without the limits, the result is the same as for the
direct regression with the same inputs, 
also when 26 variables
of the LKRv3 algorithm, including $M(\lplm)$ and $y(\lplm)$, are used as inputs. %($75.1$~GeV). 
If the
limits are applied, most of the gain is lost: $\sigma(M(\tt))=96.4$~GeV at detector level and $77$~GeV instead of
$50$~GeV at generator level. 
Thus, using the LKRv3 solution as a starting point brings no advantage, and the
direct regression is a solid standalone approach.

{\it Emulation of LKRv3.} To check the network's ability to represent the analytical algorithm,
an MLP was trained with the LKRv3 solution instead of the true values as the target. It reproduces
$M^\mathrm{LKRv3}(\tt)$ with a spread of $2.4$~GeV (median relative deviation $0.12\%$), and a network
trained in this way at generator level reproduces LKRv3 also on detector-level inputs. The main source of the better resolution
of the networks trained on the true values is their ability to learn the deviation of
the true \nunubar kinematics from the LKRv3 approximation, rather than an approximation of LKRv3.

{\it Architecture and training.} The L2 loss gives the same resolution as L1 but gives a larger bias and a less diagonal response 
(condition number $11.0$ vs.\ $8.9$). A 16 times larger MLP also does not improve
the results, and halving the training sample degrades $\sigma(M(\tt))$ by only $1$~GeV, so the
networks are not limited by the size of the MC sample. 

Overall, the resolution is limited by the information at the input, not by the model: the MLP, the
Flow, an unlimited NN-correction of LKRv3 and a larger network give equivalent results with the same
inputs. The MLP can be used as the main method, since it is about an order of magnitude cheaper than the
Flow and has a better conditioned response, while the Flow could serve as a cross-check. We leave the Flow for future
top-quark reconstruction studies, because its main potential
advantage, the per-event probability density, is not exploited by the point-estimate metrics used here.
%%%%%%%%%%%%%%%%%%%%%%%%%%%%%%%%%%%%%%%%%%%%%%%%%%%%%%%%%%%%%%%%%%%%%%
\section{Summary and outlook}
\label{sec:summary}
The kinematic reconstruction of \ttbar events in the dilepton decay channel was studied using the detector-simulated CMS MC sample available through the CMS Open Data. Performance of existing reconstruction methods was evaluated in terms of efficiency, resolution and bias. As a baseline for further development, the LKR method was chosen which provides a high reconstruction efficiency of about $98\%$ and avoids the dependence of the FKR on a fixed value of $m_t$. 

Two modifications of the LKR were developed. The first, LKRv2, applies more stringent requirements to the jet-lepton combinations used in the reconstruction and provides an improvement in the reconstruction resolution with only a marginal reduction in efficiency. The second, LKRv3, uses the approximate equality of the invariant masses and rapidities of the dilepton and neutrino-pair systems. Compared to the baseline LKR, LKRv3 improves the resolution of $M(\ttbar)$ by $7\%$, $p_T(\tt)$ by $4\%$, and $y(\tt)$ by $1\%$, while the reconstruction efficiency is reduced by about $1\%$.

Furthermore, an ML-based reconstruction was explored using the MLP and Flow methods. Both methods demonstrate similar performance. Namely, they improve the resolution of $M(\ttbar)$, $p_T(\tt)$ and $y(\tt)$ by $30\%$, $20\%$ and $15\%$, respectively, compared to LKRv3. The MLP and Flow achieve comparable reconstruction performance with the same input information. However, the MLP is approximately an order of magnitude less computationally expensive and provides a better-conditioned response matrix. This makes the MLP more practically appropriate for further applications.

These results demonstrate that ML methods can substantially improve the performance of analytical kinematic reconstruction, particularly for $M(\tt)$, which is most relevant for measurements of $m_t$. The $M(\tt)$ distribution is mostly sensitive to the top-quark mass near the production threshold at $M(\tt) \approx 2 m_t$, where currently the resolution of approximately $30$ GeV is achieved. For accurate $m_t$ extraction, it is important to simultaneously consider the effect of the quasi-bound state of the top quark and antiquark (`toponium'), however, the latter would require substantial improvement of the resolution of $M(\tt)$ in this region to $5\text{--}10$ GeV~\cite{Garzelli:2026ctb}.

The generator-level results show that a substantial part of the uncertainty in \tt kinematics originates from the intrinsic limitations of the $\nunubar$ approximation, rather than from detector resolution and/or $b$-jet assignment. In particular, the resolution of $M(\ttbar)$ and $y(\tt)$ is dominated by the $\nunubar$ approximation, while the resolution of $p_T(\tt)$ is dominated by finite detector resolution. The transfer test demonstrates that the neural network learns a genuine kinematic correction.
This indicates that the present network primarily corrects the physics-driven bias of the approximation, rather than learning detector-specific resolution effects (although detector effects still limit the achievable resolution), and could be successfully applied by another experiment, such as ATLAS. 
The achievable resolution is limited primarily by the information available in the reconstructed event and the intrinsic limitations of the neutrino-pair approximation, rather than by the capacity of the neural network.

Several directions for further development can be considered. Apart from testing alternative ML architectures to further improve the neutrino approximation, one should aim to reduce the ambiguity caused by additional jets from QCD radiation with an ML-based classifier. The methods developed in this work should also be tested on samples containing non-\ttbar events to assess their impact on background rejection and signal significance. Finally, their robustness and generality should be studied at different center-of-mass energies and, eventually, using MC simulations from ATLAS open data.

%%%%%%%%%%%%%%%%%%%%%%%%%%%%%%%%%%%%%%%%%%%%%%%%%%%%%%%%%%%%%%%%%%%%%%
\section*{Acknowledgments}
We thank the CMS Collaboration for making their data publicly available via the CERN Open Data Portal.
Support for this project was provided by the Simons Foundation and U.S. National Science Foundation cooperative agreement PHY-2323298 (IRIS-HEP).
The work of O.~Z.\ has been supported by the \emph{MSCA4Ukraine Programme} of the European Commission through the Alexander von Humboldt foundation.
We thank Olaf Behnke for interesting discussions.



%% References


\bibliographystyle{elsarticle-num}
\bibliography{refs.bib}

\end{document}